\documentstyle[aps,12pt]{revtex}
\begin{document}
\title{LARGE\ ORBITAL\ MAGNETIC\ MOMENT\\
\ IN\ FeO, FeS and FeBr$_2$ }
\author{R. J. Radwanski}
\address{C{enter for Solid State Physics, S}$^{{nt}}${\ Filip 5, 31-150 Krak%
\'{o}w,}\\
Institute of Physics, Pedagogical University, 30-084 Krak\'{o}w, {\ POLAND. }%
}
\author{Z. Ropka}
\address{{Center for Solid State Physics, S}$^{{nt}}${\ Filip 5, 31-150 Krak%
\'{o}w, POLAND.}}
\maketitle

\begin{abstract}
Magnetic moment of the Fe$^{2+}$ ion in FeO, FeS and FeBr$_2$ has been
calculated within the quantum atomistic solid-state theory to substantially
exceed a value of 4 $\mu _B$ in the magnetically-ordered state at 0 K. In
all compounds the large orbital moment, of 0.7 - 1.1 $\mu _{B\text{,}}$ has
been revealed. Our calculations show that for the meaningful analysis of
electronic and magnetic properties of FeO, FeS and FeBr$_2$ the intra-atomic
spin-orbit coupling is essentially important and that the orbital moment is
largely unquenched in 3d-ion containing compounds.


\end{abstract}


The compounds FeO, FeS and FeBr$_{2}$ have been chosen because all of them
are ionic compounds with the iron atom in the divalent state. They
crystallize in different crystallographic structures (FeO: cubic, NaCl,
a=433.4 pm; FeS: hex, NiAs, a=345 pm, c=570 pm; FeBr$_{2}$: hex, CdI$_{2}$,
a=377 pm, c=622 pm). All exhibit at low temperatures the antiferromagnetic
(AFM) ordering, but with dramatically different values of the Neel
temperature, of 198 K, 590 K and 14 K respectively. These compounds are a
subject of scientific interest for pretty long time, FeO already from the
ancient times. Despite of the simple crystallographic structure of FeO
(NaCl, B1) and the simple antiferromagnetic structure below 198 K\ its
properties are not well understood. In fact, as many of compounds containing
open shell 3d ions known as Mott insulators. Recently the problem of FeO has
been renewed by the discussion of the formation of the non-magnetic state
under megabar pressures [1-3]. Quite similar problem, high-spin to low-spin
transition, has been discussed in Ref. 4 for FeS. This experimentally-driven
discussion correlates well with the theoretical prediction that the
realization of the high-spin and low-spin states depends on the local
symmetry [5,6], but not on the strength of the cubic crystal-field
interactions. Despite of the considerations of the Co$^{3+}$ ion in Refs 5
and 6 the obtained results are directly applicable to the Fe$^{2+}$ ion as
both systems have 6 d electrons in the form of the highly-correlated 3d$^{6}$
system.

FeBr$_{2}$ exhibits the strong metamagnetic properties [7,8]. At 4.2 K an
external field of 3.15 T applied along the hexagonal axis causes a jump of
the magnetization from the almost zero value to a very big value of 4.4 $\mu
_{B}$ per the iron-ion moment. This value exceeds a theoretical value of 4.0 
$\mu _{B}$ expected for the spin-only moment. The explanation of this large
magnetic moment is a subject of the long-time debate within the 3d-magnetism
theoreticians [1-4,9,10]. A moment of 4.2 $\mu _{B}$, exceeding 4 $\mu _{B}$%
, has been also detected in FeO [2,11].

In this Letter we would like to report results of our calculations for the
magnetic moment performed within the quantum atomistic solid-state theory
(QUASST). This theory has been recently successfully applied to 3d-ion
containing compounds like NiO [12], LaCoO$_{3}$ [13,5,6] and FeBr$_{2}$
[14]. QUASST\ considers the electron correlations within the d shell as
extremely strong forming 3d$^{n}$ atomic-like electronic system. QUASST\
points out the existence of the discrete atomic-like spectrum in the 3d-ion
containing compounds associated with the quasi-atomic 3d states and the
importance of the local atomic-scale symmetry.

The QUASST\ analysis starts from the analysis of the local symmetry of the Fe%
$^{2+}$ ions in the compounds under investigations. The Fe$^{2+}$ ion in the
perfect B1 structure is in the octahedron of the oxygen ions. The
experimentally detected rhombohedral distortion [11,9] in the magnetic state
indicates on the distortion of this octahedron along the main diagonal (the
trigonal distortion). The analysis of the hexagonal FeBr$_2$\ structure
reveals the existence of the similar octahedron formed by 6 bromium atoms in
the monovalent state. The trigonal axis of the Br$_6$ octahedron is directed
along the hexagonal axis - it is the very fortunate situation that
significantly helps us in the understanding of magnetic properties of FeBr$%
_2 $ [14]. Namely, it turns out that this distortion, measured by the change
of the c/a ratio, can develop not spoiling the overall hexagonal symmetry
and what more important the magnetic moment lies along the trigonal
distortion axis. The similar octahedron is formed by the sulphur atoms in
the hexagonal FeS structure. Like in FeBr$_2$ the trigonal axis of this
octahedron is directed along the hexagonal c axis of the elementary unit
cell. Again the c/a ratio can be freely changed without breaking of the
overall symmetry. Thus, we see that in all of these compounds the Fe$^{2+}$
ions are surrounded by the octahedra formed by the negatively charged ions (O%
$^{2-}$, S$^{2-}$ and Br$^{1-}$). The size of these octahedra varies from
306 pm (FeO) to 377 pm (FeBr$_2$) and 345 pm in case of FeS. Despite of
different structures in all cases the trigonal distortion is realized.

As mentioned already the calculations have been performed within the quantum
atomistic solid-state theory, that points out the existence of the discrete
atomic-like energy spectrum. In the divalent iron state there is 6 electrons
in the open 3d shell. We consider these 6 d electrons of the Fe$^{2+}$ ion
to form the highly-correlated atomic like electronic system 3d$^6$ (note,
that in, for instance, the LDA approach these electrons are treated as
largely independent, at least at beginning). Such the system fulfills two
Hund's rules and is characterized by L=2 and S=2 [15-16]. The ground term is 
$^5$D. Its 25-fold degeneracy is removed by the (dominant) cubic crystal
field (CEF)\ interactions (cubic CEF parameter B$_4$ = +200 K, the
octahedral site), the intra-atomic spin-orbit coupling ($\lambda _{s-o}$ =
-150 K) and an off-cubic trigonal distortion (parameter B$_2^0$ of -30 K has
been derived in FeBr$_2$). These parameters corresponds to the overall
energies: $\Delta _{CF}^{oc}=2$ eV, $\Delta _{s-o}^{}=0.13$ eV and $\Delta
_t=0.03$ eV.

These parameters yield the doublet ground state, that is highly magnetic.
Quite close to the ground doublet is a non-magnetic singlet. Such the
triplet structure is characteristic for the Fe$^{2+}$ and Co$^{3+}$ ions (3d$%
^6$ systems, also 4d$^6$ and 5d$^6$ systems) in the octahedral crystal field
[15]. The triplet is split by the off-cubic distortion [5,6]. The
non-magnetic singlet becomes the ground state in LaCoO$_3$. The magnetism of
FeO, FeS and FeBr$_2$ develops on the magnetic doublet when the intersite
spin-dependent interactions are switched on in the self-consistent way.
These spin-dependent interactions, approximated by the molecular-field
approach, set up the internal molecular field of 116, 430 and 5.1 T,
respectively that splits the ground-state doublet. We would like to point
out that the magnetism (the value and the direction of the magnetic moment,
the temperature dependence of the magnetic moment) and the low-energy, 
%TCIMACRO{\TEXTsymbol{<} }%
%BeginExpansion
\mbox{$<$}
%EndExpansion
4 eV, electronic structure all of these compounds is understood
consistently. The higher Neel temperature - the larger internal field and
the larger splitting of the ground-state doublet is. One should note that
the involved magnetic energies are of order of 0.02-0.03 eV (FeO) only, i.e.
they are relatively small.

The calculated atomic-like magnetic moment of the Fe$^{2+}$ ion in the ionic
FeBr$_2$ compound in the magnetically-ordered state at 0 K turns out to be
4.32 $\mu _B$ [14]. This moment is composed from the spin moment of +3.52 $%
\mu _B$ and the orbital moment of +0.80 $\mu _B$. In the paramagnetic state
the total moment amounts to 4.26 $\mu _B$ (3.48 and 0.78 $\mu _B$). In FeO
at 0 K the spin and orbital moments amount to 3.70 and 0.90 $\mu _B$ (total:
4.60 $\mu _B$), and in FeS to 3.94 and 1.06 $\mu _B$ (total: 5.00 $\mu _B$),
respectively. The orbital moment adds to the spin moment as one could expect
knowing the 3$^{rd}$ Hund's rule.

The crystal-field interactions in the hexagonal structure of FeBr$_{2}$ and
FeS cause the Fe-ion moment to lie along the hexagonal axis. The hexagonal
axis is the trigonal axis of the local Br$^{1-}$ and S$^{2-}$ octahedra. As
a consequence the antiferromagnet FeBr$_{2}$ is very good example of the
Ising-type magnetic compound. The metamagnetic transition is associated with
the reversal of the AFM moment and it occurs at 3.15 T for FeBr$_{2}$. We
expect the metamagnetic transition to occur in FeO at about 75 T for fields
applied along the 
%TCIMACRO{\TEXTsymbol{<}}%
%BeginExpansion
\mbox{$<$}%
%EndExpansion
111%
%TCIMACRO{\TEXTsymbol{>} }%
%BeginExpansion
\mbox{$>$}
%EndExpansion
direction. An enormous field of 260 T is needed for the metamagnetic
transition in FeS (along the hexagonal c axis).

The Fe$^{2+}$-ion moment strongly depends on the local symmetry. The local
symmetry is reflected in the symmetry of the CEF. In case of the octahedral
crystal field formed by the negative charges the elongation of the
octahedron along the diagonal (than the rhombohedral angle $\alpha _R$ 
%TCIMACRO{\TEXTsymbol{<}}%
%BeginExpansion
\mbox{$<$}%
%EndExpansion
60$^o$ like it is observed in FeO) pushes the magnetic doublet down and
helps the magnetism, the stretching can lead to the non-magnetic singlet
ground state like in LaCoO$_3$. According to us the stretching of the local
octahedron is also the reason for the low-spin state in diamagnetic Fe$^{2+}$%
-ion compounds, but the further discussion of this problem goes beyond the
present Letter.

Finally we would like to note that our atomic-like theory is, in fact, no
free parameter theory. All parameters have the clear physical meaning. The
spin-orbit coupling is known from the atomic physics. The parameter B$_4^0$
can be quite well estimated from conventional charge multipolar
interactions- the sign, for instance, is fully determined by the octahedral
anion surrounding [15,16]. The parameter B$_4^0$ is expected to decrease
going from FeO to FeBr$_2$ via FeS (the iron--anion distance increases) - we
took for the present consideration the constant value for the simplicity
reason. It turns out that the size of B$_4^0$ does not affect much the
ground-state results as the B$_4^0$ energy scale, of 1-2 eV, is much larger
than the spin-orbit, distortion and magnetic energy scales. Moreover,
preceding some critics we should say that we start the considerations from
the iron atom in the octahedral anion surrounding but due to the
translational symmetry the whole crystal is built up from such corner, edge
or wall sharing octahedra.

In conclusion, the Fe$^{2+}$ ion in FeO, FeS and FeBr$_2$ exhibits large
atomic-like magnetic moment, 
%TCIMACRO{\TEXTsymbol{>} }%
%BeginExpansion
\mbox{$>$}
%EndExpansion
4 $\mu _B,$ with the very substantial orbital magnetic moment (0.7-1.1 $\mu
_B$). Such large orbital moment indicates that the orbital moment has to be
''unquenched'' in the modern solid-state theory of 3d-ion containing
compounds.

E-mail (R.J.Radwanski) - sfradwan@cyf-kr.edu.pl


\begin{references}
\bibitem{1} J.Badro, V.V.Struzhkin, J.Shu, R.J.Hemley, H.K.Mao, C.-C Kao,
J.-P.Rueff and G.Shen, Phys.Rev.Lett. 83 (1999) 4101.

\bibitem{2} Z.Fang, I.V.Solovyev, H.Sawada and K.Terakura, Phys.Rev. B 59
(1999) 762.

\bibitem{3} Z.Fang, K.Terakura, H.Sawada, T.Miyazaki and I.V.Solovyev,
Phys.Rev.Lett. 81 (1998) 1027.

\bibitem{4} J.-P.Rueff, C.-C.Kao, V.V.Struzhkin, J.Badro, J.Shu, R.J.Hemley
and H.K.Mao, Phys.Rev.Lett. 82 (1999) 3284.

\bibitem{5} R.J.Radwanski and Z.Ropka, Solid State Commun. 112 (1999) 621.

\bibitem{6} R.J.Radwanski and Z.Ropka, Local off-cubic distortion the cause
for the low- and high-spin states of the Co$^{3+}$ ion,
http://xxx.lanl.gov/abs/cond-mat/0006269.

\bibitem{7} M.K.Wilkinson, J.W.Cable, E.O.Wollan and W.C.Koehler, Phys.Rev.
113 (1959) 497.

\bibitem{8} K.Katsumata, H.Aruga Katori, S.M.Shapiro and G.Shirane,
Phys.Rev. B 55 (1997) 11466.

\bibitem{9} D.G.Isaak, R.E.Cohen, M.J.Mehl and D.J.Singh, Phys. Rev. B 47
(1993) 7720.

\bibitem{10} R.J.Birgenau, W.B.Yelon, E.Cohen and J.Makovsky, Phys.Rev. B 5
(1972) 2607.

\bibitem{11} P.D.Battle and A.K.Cheetham, J.Phys. C 12 (1979) 337.

\bibitem{12} R.J.Radwanski, Z.Ropka and R.Michalski, The orbital moment in
NiO, http://xxx.lanl.gov/abs/cond-mat/0005358.

\bibitem{13} R.J.Radwanski and Z.Ropka, Physica B 281\&282 (2000) 507.

\bibitem{14} Z.Ropka, R.Michalski and R.J.Radwanski, Electronic and magnetic
properties of FeBr$_2$, http://xxx.lanl.gov/abs/cond-mat/0005502.

\bibitem{15} R.J.Radwanski and Z.Ropka, Relativistic effects in the
electronic structure for the 3d paramagnetic ions,
http://xxx.lanl.gov/abs/cond-mat/9907140.

\bibitem{16} A.Abragam and B.Bleaney, in: Electron Paramagnetic Resonance of
Transition Ions, (Clarendon Press, Oxford) 1970. ch. 7 (the intermediate
crystal-field approach).
\end{references}
\end{document}